

\def\singlespace{\baselineskip 12pt \lineskip 1pt \parskip 2pt plus 1 pt}


\singlespace
\magnification=\magstep1


\raggedbottom

\def\refto#1{$^{#1}$}           
\def\ref#1{ref.~#1}                     
\def\Ref#1{#1}                          
\gdef\refis#1{\item{#1.\ }}                     
\def\beginparmode{\endmode
  \begingroup \def\endmode{\par\endgroup}}
\let\endmode=\par
\def\body{\beginparmode}
\def\head#1{                    
  \goodbreak\vskip 0.5truein    
  {\centerline{\bf{#1}}\par}
   \nobreak\vskip 0.25truein\nobreak}
\def\references                 
Phys Rev
  {\head{References}            
   \beginparmode
   \frenchspacing \parindent=0pt \leftskip=1truecm
   \parskip=8pt plus 3pt \everypar{\hangindent=\parindent}}
\def\endreferences{\body}

\font\lgh=cmbx10 scaled \magstep2

\def\hb{\hfil\break}

\def\ale{\mathrel{\hbox{\rlap{\hbox{\lower4pt\hbox{$\sim$}}}\hbox{$<$}}}}
\def\age{\mathrel{\hbox{\rlap{\hbox{\lower4pt\hbox{$\sim$}}}\hbox{$>$}}}}

\catcode`@=11
\newcount\r@fcount \r@fcount=0
\newcount\r@fcurr
\immediate\newwrite\reffile
\newif\ifr@ffile\r@ffilefalse
\def\w@rnwrite#1{\ifr@ffile\immediate\write\reffile{#1}\fi\message{#1}}

\def\writer@f#1>>{}
\def\referencefile{
  \r@ffiletrue\immediate\openout\reffile=\jobname.ref%
  \def\writer@f##1>>{\ifr@ffile\immediate\write\reffile%
    {\noexpand\refis{##1} = \csname r@fnum##1\endcsname = %
     \expandafter\expandafter\expandafter\strip@t\expandafter%
     \meaning\csname r@ftext\csname r@fnum##1\endcsname\endcsname}\fi}%
  \def\strip@t##1>>{}}

\def\citeall#1{\xdef#1##1{#1{\noexpand\cite{##1}}}}
\def\cite#1{\each@rg\citer@nge{#1}}	

\def\each@rg#1#2{{\let\thecsname=#1\expandafter\first@rg#2,\end,}}
\def\first@rg#1,{\thecsname{#1}\apply@rg}	
\def\apply@rg#1,{\ifx\end#1\let\next=\relax
\else,\thecsname{#1}\let\next=\apply@rg\fi\next}

\def\citer@nge#1{\citedor@nge#1-\end-}	
\def\citer@ngeat#1\end-{#1}
\def\citedor@nge#1-#2-{\ifx\end#2\r@featspace#1 
  \else\citel@@p{#1}{#2}\citer@ngeat\fi}	
\def\citel@@p#1#2{\ifnum#1>#2{\errmessage{Reference range #1-#2\space is bad.}%
    \errhelp{If you cite a series of references by the notation M-N, then M and
    N must be integers, and N must be greater than or equal to M.}}\else%
 {\count0=#1\count1=#2\advance\count1 by1\relax\expandafter\r@fcite\the\count0,%
  \loop\advance\count0 by1\relax
    \ifnum\count0<\count1,\expandafter\r@fcite\the\count0,%
  \repeat}\fi}

\def\r@featspace#1#2 {\r@fcite#1#2,}	
\def\r@fcite#1,{\ifuncit@d{#1}
    \newr@f{#1}%
    \expandafter\gdef\csname r@ftext\number\r@fcount\endcsname%
                     {\message{Reference #1 to be supplied.}%
                      \writer@f#1>>#1 to be supplied.\par}%
 \fi%
 \csname r@fnum#1\endcsname}
\def\ifuncit@d#1{\expandafter\ifx\csname r@fnum#1\endcsname\relax}%
\def\newr@f#1{\global\advance\r@fcount by1%
    \expandafter\xdef\csname r@fnum#1\endcsname{\number\r@fcount}}

\let\r@fis=\refis			
\def\refis#1#2#3\par{\ifuncit@d{#1}
   \newr@f{#1}%
   \w@rnwrite{Reference #1=\number\r@fcount\space is not cited up to now.}\fi%
  \expandafter\gdef\csname r@ftext\csname r@fnum#1\endcsname\endcsname%
  {\writer@f#1>>#2#3\par}}

\def\ignoreuncited{
   \def\refis##1##2##3\par{\ifuncit@d{##1}%
     \else\expandafter\gdef\csname r@ftext\csname r@fnum##1\endcsname\endcsname%
     {\writer@f##1>>##2##3\par}\fi}}

\def\r@ferr{\endreferences\errmessage{I was expecting to see
\noexpand\endreferences before now;  I have inserted it here.}}
\let\r@ferences=\references
\def\references{\r@ferences\def\endmode{\r@ferr\par\endgroup}}

\let\endr@ferences=\endreferences
\def\endreferences{\r@fcurr=0
  {\loop\ifnum\r@fcurr<\r@fcount
    \advance\r@fcurr by 1\relax\expandafter\r@fis\expandafter{\number\r@fcurr}%
    \csname r@ftext\number\r@fcurr\endcsname%
  \repeat}\gdef\r@ferr{}\endr@ferences}


\let\r@fend=\endpaper\gdef\endpaper{\ifr@ffile
\immediate\write16{Cross References written on []\jobname.REF.}\fi\r@fend}

\catcode`@=12

\citeall\refto		
\citeall\ref		%
\citeall\Ref		%

\def\singlespace{\baselineskip 12pt \lineskip 1pt \parskip 2pt plus 1 pt}

\def\today{\number\day\enspace
     \ifcase\month\or January\or Febuary\or March\or April\or May\or
     June\or July\or August\or September\or October\or
     November\or December\fi \enspace\number\year}
\def\clock{\count0=\time \divide\count0 by 60
    \count1=\count0 \multiply\count1 by -60 \advance\count1 by \time
    \number\count0:\ifnum\count1<10{0\number\count1}\else\number\count1\fi}
\footline={\hss -- \folio\ -- \hss}

\def\deg{\ifmmode^\circ\else$^\circ$\fi}
\def\solar{\ifmmode_{\mathord\odot}\else$_{\mathord\odot}$\fi}
\def\jref#1 #2 #3 #4 {{\par\noindent \hangindent=3em \hangafter=1 
      \advance \rightskip by 5em #1, {\it#2}, {\bf#3}, #4.\par}}
\def\ref#1{{\par\noindent \hangindent=3em \hangafter=1 
      \advance \rightskip by 5em #1.\par}}
\newcount\eqnum
\def\nexteq{\global\advance\eqnum by1 \eqno(\number\eqnum)}
\def\lasteq#1{\if)#1[\number\eqnum]\else(\number\eqnum)\fi#1}
\def\preveq#1#2{{\advance\eqnum by-#1
    \if)#2[\number\eqnum]\else(\number\eqnum)\fi}#2}

\def\tableheight{\vrule width 0pt height 8.5pt depth 3.5pt}
{\catcode`|=\active \catcode`&=\active 
    \gdef\tabledelim{\catcode`|=\active \let|=\vbar
                     \catcode`&=\active \let&=\nobar} }
\def\table{\begingroup
    \def\twidth{\hsize}
    \def\tablewidth##1{\def\twidth{##1}}
    \def\defaultheight{\vrule width 0pt height 8.5pt depth 3.5pt}
    \def\heightdepth##1{\dimen0=##1
        \ifdim\dimen0>5pt 
            \divide\dimen0 by 2 \advance\dimen0 by 2.5pt
            \dimen1=\dimen0 \advance\dimen1 by -5pt
            \vrule width 0pt height \the\dimen0  depth \the\dimen1
        \else  \divide\dimen0 by 2
            \vrule width 0pt height \the\dimen0  depth \the\dimen0 \fi}
    \def\spacing##1{\def\defaultheight{\heightdepth{##1}}}
    \def\nextheight##1{\noalign{\gdef\tableheight{\heightdepth{##1}}}}
    \def\end{\cr\noalign{\gdef\tableheight{\defaultheight}}}
    \def\zerowidth##1{\omit\hidewidth ##1 \hidewidth}    
    \def\hline{\noalign{\hrule}}
    \def\skip##1{\noalign{\vskip##1}}
    \def\bskip##1{\noalign{\hbox to \twidth{\vrule height##1 depth 0pt \hfil
        \vrule height##1 depth 0pt}}}
    \def\header##1{\noalign{\hbox to \twidth{\hfil ##1 \unskip\hfil}}}
    \def\bheader##1{\noalign{\hbox to \twidth{\vrule\hfil ##1 
        \unskip\hfil\vrule}}}
    \def\spanloop{\span\omit \advance\mscount by -1}
    \def\extend##1##2{\omit
        \mscount=##1 \multiply\mscount by 2 \advance\mscount by -1
        \loop\ifnum\mscount>1 \spanloop\repeat \ \hfil ##2 \unskip\hfil}
    \def\vbar{&\vrule&}
    \def\nobar{&&}
    \def\hdash##1{ \noalign{ \relax \gdef\tableheight{\heightdepth{0pt}}
        \toks0={} \count0=1 \count1=0 \putout##1\end 
        \toks0=\expandafter{\the\toks0 &\end} \xdef\piggy{\the\toks0} }
        \piggy}
    \let\e=\expandafter
    \def\putspace{\ifnum\count0>1 \advance\count0 by -1
        \toks0=\e\e\e{\the\e\toks0\e&\e\multispan\e{\the\count0}\hfill} 
        \fi \count0=0 }
    \def\putrule{\ifnum\count1>0 \advance\count1 by 1
        \toks0=\e\e\e{\the\e\toks0\e&\e\multispan\e{\the\count1}\leaders\hrule\hfill}
        \fi \count1=0 }
    \def\putout##1{\ifx##1\end \putspace \putrule \let\next=\relax 
        \else \let\next=\putout
            \ifx##1- \advance\count1 by 2 \putspace
            \else    \advance\count0 by 2 \putrule \fi \fi \next}   }
\def\tablespec#1{
    \def\vdimens{\noexpand\tableheight}
    \def\tabby{\tabskip=0pt plus100pt minus100pt}
    \def\r{&################\tabby&\hfil################\unskip}
    \def\c{&################\tabby&\hfil################\unskip\hfil}
    \def\l{&################\tabby&################\unskip\hfil}
    \edef\templ{\noexpand\vdimens ########\unskip  #1 
         \unskip&########\tabskip=0pt&########\cr}
    \tabledelim
    \edef\body##1{ \vbox{
        \tabskip=0pt \offinterlineskip
        \halign to \twidth {\templ ##1}}} }

\newbox\grsign \setbox\grsign=\hbox{$>$}
\newdimen\grdimen \grdimen=\ht\grsign
\newbox\laxbox \newbox\gaxbox
\setbox\gaxbox=\hbox{\raise.5ex\hbox{$>$}\llap
	{\lower.5ex\hbox{$\sim$}}}\ht1=\grdimen\dp1=0pt
\setbox\laxbox=\hbox{\raise.5ex\hbox{$<$}\llap
	{\lower.5ex\hbox{$\sim$}}}\ht2=\grdimen\dp2=0pt
\def\simlt{\mathrel{\copy\laxbox}}

\def\uJy{\ifmmode{\,\mu{\rm Jy}}\else$\,{\mu{\rm Jy}}$\fi}
\def\mJy{\ifmmode{\,{\rm mJy}}\else${\,{\rm mJy}}$\fi}
\def\MHz{\ifmmode{\,{\rm MHz}}\else{$\,{\rm MHz}$}\fi}
\def\GHz{\ifmmode{\,{\rm GHz}}\else{$\,{\rm GHz}$}\fi}
\def\solar{\ifmmode_{\mathord\odot}\else$_{\mathord\odot}$\fi}
\def\Msolar{\ifmmode{\, {\rm M\solar}}\else{${\, {\rm M\solar}}$}\fi}
\def\Rsolar{\ifmmode{\, {\rm R\solar}}\else{${\, {\rm R\solar}}$}\fi}
\def\kms{\ifmmode{\,{\rm km\,s^{-1}}}\else${\,{\rm km\,s^{-1}}}$\fi}
\def\kpc{\ifmmode{\,{\rm kpc}}\else${\,{\rm kpc}}$\fi}
\def\us{\ifmmode{\,\mu{\rm s}}\else$\,{\mu{\rm s}}$\fi}
\def\ms{\ifmmode{\,{\rm ms}}\else$\,{{\rm ms}}$\fi}
\def\y{\ifmmode{\,{\rm y}}\else$\,{\rm y}$\fi}
\def\h{\ifmmode{^{\rm h}}\else$^{\rm h}$\fi}
\def\m{\ifmmode{^{\rm m}}\else$^{\rm m}$\fi}
\def\s{\ifmmode{^{\rm s}}\else$^{\rm s}$\fi}
\def\Lmin{\ifmmode{L_{min}}\else{$L_{min}$}\fi}

\input psfig.sty


\def\Image{1}
\def\Light-Curve{2}
\def\Spectrum{3}

\def\TableMagnitudes{1}


\hrule
\bigskip
\line{\lgh The unusual afterglow of GRB 980326: evidence for \hb}
\line{\lgh the gamma-ray burst/supernova connection\hb}
\bigskip

\def\Palomar  {$^1$}
\def\NRAO     {$^2$}
\def\UCB      {$^3$}
\def\JHU      {$^4$}
\def\IGPP     {$^5$}
\def\CPA      {$^6$}
\def\LBL      {$^7$}
\def\ESO      {$^8$}
\def\IAS      {$^9$}

\line{      J. S. Bloom\Palomar,             
            S. R. Kulkarni\Palomar,          
            S. G. Djorgovski\Palomar,        
            A. C. Eichelberger\Palomar,      
\hb}
\line{
            P.    C\^ot\'e\Palomar,          
            J. P. Blakeslee\Palomar,         
            S. C. Odewahn\Palomar,           
	    F. A. Harrison\Palomar,          
            D. A. Frail\NRAO,	             
\hb}
\line{
            A. V. Filippenko\UCB,            
            D. C. Leonard\UCB,               
            A. G. Riess\UCB,                 
            H. Spinrad\UCB,                  
\hb}
\line{
            D. Stern\UCB,                    
            A. Bunker\UCB,                   
            A. Dey\JHU,                      
            S. A. Stanford\IGPP,             
            B. Grossan\CPA,                  
\hb}
\line{
            S. Perlmutter\LBL,               
            R. A. Knop\LBL,                  
            I. M. Hook\ESO,                  
            \&\
            M.    Feroci\IAS                
\hb}

\medskip

\line{\Palomar Palomar Observatory 105-24, Caltech, Pasadena, CA 91125,
USA\hb}

\line{\NRAO National Radio Astronomy Observatory, P. O. Box O,
       Socorro, NM 87801, USA\hb}

\line{\UCB Department of Astronomy, University of California, Berkeley,
CA 94720-3411 USA\hb}


\line{\JHU National Optical Astronomy Observatories, 950 N. Cherry,
Ave. \hb}
\line{Tucson, AZ 85719, USA\hb}

\line{\IGPP Institute of Geophysics and Planetary Physics,
Lawrence\hb}
\line{Livermore National Laboratory, 7000 East Avenue, P. O. Box 808,
L-413,\hb}
\line{Livermore, CA 94551-9900, USA\hb}

\line{\CPA Center for Particle Astrophysics, University of California,
Berkeley, CA 94720 USA}

\line{\LBL Lawrence Berkeley National Laboratory, Berkeley, CA 94720,
USA\hb}

\line{\ESO European Southern Observatory, D-85748 Garching, Germany\hb}

\line{\IAS Istituto di Astrofisica Spaziale, CNR,
           via Fosso del Cavaliere, Roma I-00133, Italy\hb}

\medskip
\hrule
\bigskip

\noindent{\it  This manuscript was submitted to Nature on March 23,
1999.  We are making this m.s. available on astroph given the rapid
progress in the field of GRBs.  You are free to refer to this paper in
your own paper. However, we do place restrictions on any dissemination
in the popular media. The article is under embargo until it is
published. For further enquiries, please contact Shri Kulkarni ({\tt
srk@astro.caltech.edu}) or Joshua Bloom ({\tt
jsb@astro.caltech.edu}).}

\bigskip

\noindent{\bf Cosmic gamma-ray bursts (GRBs) have been firmly
established as one of the most powerful phenomena in the Universe,
releasing electromagnetic energy approaching the rest-mass energy of a
neutron star in a few seconds. The two currently popular models for
GRB progenitors are the coalescence of two compact objects (such as
neutron stars or black holes) or collapse of a massive star.  An
unavoidable consequence of the latter model is that a bright
supernovae should accompany the GRB.  The emission from this supernova
competes with the much brighter afterglow produced by the relativistic
shock that gives rise to the GRB itself.  Here we present evidence for
an unusual light curve for GRB 980326 based on new optical
observations.  The transient brightened $\sim 3$ weeks after the burst
to a flux sixty times larger than that extrapolated from the rapid
decay seen at early time.  Furthermore, the spectrum changed
dramatically and became extremely red. We argue that the new source is
the underlying supernova.  If our hypothesis is true then this would
be the first evidence for a supernova connection with GRBs at
cosmological distances. We suggest that GRBs with long durations are
associated with death of massive stars.}

\bigskip
\bigskip


The origin of GRBs remained elusive for a period of nearly three
decades after their discovery\refto{Klebesadel73}.  Beginning in 1997,
however, the prompt localization of GRBs by the Italian-Dutch
satellite BeppoSAX\refto{Boella97} and the All Sky
Monitor\refto{Levine96} on board the X-ray Timing Explorer led to the
discovery of the GRB afterglow phenomenon -- emission at lower
energies: X-ray\refto{Costa97}, optical\refto{JvP97}, and
radio\refto{Frail97}.

\refis{Klebesadel73}
        Klebesadel, R. W., Strong, I. B., \&\ Olson, R. A.
        Observations of gamma-ray bursts of cosmic origin.  
        {\it Astrophys. J.} {\bf 182}, L85--L88 (1973).

\refis{Boella97}
	Boella, G.  et al.
	BeppoSAX, the wide band mission for x-ray astronomy.
	{\it Astron. Astrophys. Suppl. Ser.} {\bf 122}, 299--399 (1997).
	
\refis{Levine96}
	Levine, A. M., Bradt, H., Cui, W., Jernigan, J. G., 
	Morgan, E. H., Remillard, R., Shirey, R. E. \&\ Smith, D. A.
	First Results from the All-Sky Monitor on the Rossi X-ray
	Timing Explorer.
	{\it Astrophys. J.} {\bf 469}, L33-L36 (1996).

\refis{Costa97} Costa, E. et al.
            Discovery of an X-ray afterglow associated with the gamma-ray
            burst of 28 February 1997.
            {\it Nature} {\bf 387}, 783--785 (1997).

\refis{JvP97}
	van Paradijs, J.  et al.
 	Transient optical emission from the error box of the 
	$\gamma$-ray burst of 28 February 1997.
 	{\it Nature} {\bf 386} 686--689 (1997).

\refis{Frail97}
	Frail, D. A., Kulkarni, S. R., Nicastro, L., Feroci, M. 
	\&\ Taylor, G.  B.
	The radio afterglow from the gamma-ray burst of 8 May 1997.
	{\it Nature} {\bf 389}, 261--263 (1997).

The persistence of the afterglow emission (days at X-ray wavelengths,
weeks to months at optical wavelengths, months to a year at radio
wavelengths) enabled astronomers to carry out detailed observations
which led to fundamental advances in our understanding of these
sources: (1) the demonstration that GRBs are at cosmological
distances\refto{Metzger97}; (2) the proof that these sources expand
with relativistic speeds\refto{Frail97}; and (3) the realization that
the electromagnetic energy released in these objects exceeds that in
supernovae\refto{WKF98} and, in some cases, the released energy is
comparable to the rest mass energy of a neutron
star\refto{Kulkarni98a,Djorgovski98a, Kulkarni99,Anderson99}.

\refis{Metzger97}   
        Metzger, M. R., Djorgovski, S. G., Kulkarni, S. R.,
        Steidel, C. C., Adelberger, K. L., Frail, D. A.,
        Costa, E. \&\ Frontera, F. 
        Spectral Constraints on the redshift of the optical
        counterpart to the gamma-ray burst of May 8, 1997.
        {\it Nature} {\bf 387}, 878--879 (1997). 

\refis{WKF98}
	Waxman, E., Kulkarni, S. R. \&\ Frail, D. A. 
	Implications of the Radio Afterglow from the Gamma-ray
	Burst of 1997 May 8.
	{\it Astrophys. J.} {\bf 502}, L119--L122 (1998).

\refis{Kulkarni98a}
        Kulkarni, S.~R. et~al.
        Identification of a host galaxy at redshift $z$=3.42 for the
        $\gamma$-ray burst of 14 Dec 1997.
        {\it Nature} {\bf 393}, 35--39 (1998).

\refis{Djorgovski98a}
        Djorgovski, S. G., Kulkarni, S. R., Bloom, 
        J. S., Goodrich, R., Frail, D. A., Piro, L., \&\ Palazzi, E.
        Spectroscopy of the Host Galaxy of the Gamma--Ray Burst 980703.
	Astrophys. J. {\bf 508}, L17--L20 (1998).

\refis{Kulkarni99}
	Kulkarni, S. R. et al.
	The afterglow, the redshift, and the extreme energetics of the
	gamma-ray burst 990123.  
	{\it Nature} {\bf 398}, 389--394 (1999).

\refis{Anderson99}
	Andersen, M. I. et al. 
	Spectroscopic limits on the distance and energy release of
	GRB 990123.
	{\it Science} {\bf 283}, 2075--2077 (1999).

Despite these advances, we are still largely in the dark about the
nature of the GRB progenitors. Though there are a number of models for
their origin, the currently popular models involve the formation of
black holes resulting from either the coalescence of neutron
stars\refto{Paczynski86,Goodman86,Narayan92} or the death of massive
stars\refto{Woosley93,Paczynski98}.  The small offsets of GRBs with
respect to their host galaxies and the association of GRBs with dusty
regions and star-formation regions favors the latter, the so-called
hypernova scenario\refto{Paczynski98}.  However, this evidence is
indirect and also limited by the small number of well-studied GRBs.

\refis{Paczynski86}
     	Paczy\`nski, B.
     	Gamma-ray bursters at cosmological distances.
    	{\it Astrophys. J.}, {\bf 308}, L43--L46 (1986).

\refis{Goodman86}
        Goodman, J.
	Are Gamma-Ray Bursts Optically Thick?
	{\it Astrophys. J.}, {\bf 308}, L47--L50 (1986).

\refis{Narayan92}
        Narayan, R., Paczy\`nski, B., Piran, T.
	Gamma-ray bursts as the death throes of massive binary stars.
       {\it Astrophys. J.}, {\bf 95}, L83--L86 (1992).

\refis{Woosley93}
	Woosley, S. E. 
	Gamma-Ray Bursts from Stellar Collapse to a Black Hole?
	{\it Astrophys. J.}, {\bf 405}, 273--277 (1993).

\refis{Paczynski98}
	Paczyn\'ski, B.
	Are Gamma-Ray bursts in Star-Forming Regions?
	{\it Astrophys. J.} {\bf 494}, L45--L48 (1998).

The most direct evidence for a massive star origin would be the
observation of a supernova coincident with a GRB.  Here we present
observations of GRB 980326 and argue for the presence of such an
underlying supernova. If our conclusions are correct then the
implication is that at least some fraction of GRBs, perhaps the entire
class of long duration GRBs, represent the end point of the most
massive stars. Furthermore, if the
association\refto{Galama98,Kulkarni98b} of GRB 980425 with a bright
supernova in a nearby galaxy holds, then the apparent $\gamma$-ray
luminosity of GRBs ranges over six orders of magnitude.

\refis{Galama98} 
	Galama, T. et al. 1998,
	An unusual supernova in the error box of the gamma-ray burst of
	25 April 1998.
	{\it Nature} {\bf 395}, 670-672.

\refis{Kulkarni98b}
	Kulkarni, S. R. et al. 1998,
	Radio emission from the unusual supernova 1998bw and its
	association with the $\gamma$-ray burst of 25 April 1998.
	{\it Nature} {\bf 395}, 663--669.

\medskip
\noindent{\bf The unusual optical afterglow}
\medskip

Following the localization of GRB 980326 by BeppoSAX (ref.
\Ref{Cel98}), Groot et al.\refto{Groot98} quickly identified the
optical afterglow.  Our optical follow-up program began at the Keck
Observatory, just 10 hr after the burst.  A log of these observations
is given in Table \TableMagnitudes.

\refis{Cel98}
	Celidonio, G., Coletta, A., Feroci, M., Piro, L., Soffitta,
	P., in 't Zand, J., Muller, J., Palazzi, E.
	GRB 980326.
	{\it I.A.U.C.} {\bf 6851} (1998).

\refis{Groot98}
	Groot, P. J. et al.
	The Rapid Decay of the Optical Emission from GRB 980326 and
	Its Possible Implications.
	{\it Astrophys. J.} {\bf 502}, L123--127 (1998).

In Figure~\Light-Curve we present our $R$-band photometry along with
those reported in the literature.  Restricting to data taken within
the first month of the burst reported in the
literature\refto{Groot98,Valdes98} and via the ``GRB Coordinates
Network" (GCN) (ref. \Ref{GCN}), we find a characteristic power law
decay in the flux versus time followed followed by an apparent
flattening.  The usual interpretation is that the decaying flux is the
afterglow emission, while the constant flux reflects the presence of
the host galaxy.  Indeed, earlier\refto{gcn57} we attributed the
entire observed flux on April 17th to the host galaxy.

\refis{Valdes98}
	Valdes, F., Jannuzi, B., \&\ Rhoads, J.
	 GRB980326, optical observations.
	 {\it G.C.N.} {\bf 156}, (1998).

\refis{GCN}
	The GRB Coordinates Network.
	http://gcn.gsfc.nasa.gov/gcn/

\refis{gcn57}
	Djorgovski, S. G., Kulkarni, S. R., C\^ ot\' e, P., Blakeslee, J.
	Bloom, J. S., Odewahn, S. C.
	GRB980326, Optical Observations.
	{\it G.C.N.} {\bf 189}, (1999).

However, to our surprise, our more recent observations (first carried
out nine months after the GRB event) showed no galaxy at the position
of the optical transient (OT); see Figure~\Image.  We estimate a
2-$\sigma$ upper limit of $R>27.3$ magnitude (see Table
\TableMagnitudes). This is almost a factor of 10 less flux than that
reported from our April 17th detection.  A secure conclusion is that
the presumed host galaxy of GRB 980326, assuming the GRB was
coincident with the host (as appears to be the case for all other
well-studied GRBs to date) is fainter than $R \sim 27$ magnitude.
This conclusion is not alarming since such faint (or fainter) galaxies
are indeed expected from studies\refto{Mao98,FruHogg99} of the
properties of cosmological GRB host galaxies.

\refis{Mao98}
	Mao, S. \&\ Mo, H. J.
	The nature of the host galaxies for gamma-ray bursts.
	{\it Astron. Astrophys.} {\bf 339}, L1--L4 (1998).
\refis{FruHogg99}
	Hogg, D. W., Fruchter, A. S.
	The faint galaxy hosts of Gamma-Ray bursts.
	{\it Astrophys. J.} in press (1999).

Having established that the host galaxy of GRB 980326 is faint, we are
forced to conclude that the OT did not continue the rapid decay
it exhibited initially.  Instead, we find two phases of the light curve
(Figure \Light-Curve): a steeply declining initial
phase ($t\simlt 5$ d) and a
subsequent
rebrightening phase ($t\sim $ 3--4 weeks). Following the rebrightening,
the source
appears to have faded away to an undetectable level by the time
of our next observation (9 months after the burst).

In previously studied bursts the optical afterglow emission has been
modeled by a power law function, flux $\propto t^{\alpha}$; here, $t$
is time since the burst and $\alpha$ the power law index.  In some
bursts, at early times ($t$ less than a day or so), significant
deviations have been seen, e.g., GRB 970508 (ref.~\Ref{Djo97}).
At late
times, in some bursts, deviations manifest as steepening
(i.e. $\alpha$ becoming smaller) of light curves, e.g., GRB 990123
(ref.~\Ref{Kulkarni99}).

\refis{Djo97}
	Djorgovski, S. G. et al.
	The optical counterpart to the gamma-ray burst GRB 970508
	{\it Nature}, {\bf 387}, 876--878 (1997).

It is against this backdrop of the observed
afterglow phenomenology that  we now analyze the light curve in Figure
\Light-Curve.  The declining phase cannot be fit by a simple power law
($\chi^2=72$ for 9 degrees of freedom). From Figure \Light-Curve\ it
is clear that the flux already starts flattening by day 3.
Restricting the analysis to the first two days, we obtain $\alpha =
-2.0 \pm 0.1$, consistent with previous analysis\refto{Groot98}.

Such power law decays are usually interpreted as arising from
electrons shocked by the explosive debris sweeping up the ambient
medium\refto{Katz94,MR97,Vietri97,Waxman97a}.  Assuming that the
electrons behind the shock are accelerated to a power-law differential
energy distribution with index $-p$, on general grounds\refto{Sari98}
we expect that the afterglow flux, $f_\nu(t)\propto
t^{\alpha}\nu^{\beta}$; here $f_{\nu}(t)$ is the flux at frequency
$\nu$ and time $t$.  The value of $\alpha$ and $\beta$ depend on $p$,
the geometry of the emitting surface\refto{Meszaros98} (spherical
versus collimation) and the radial distribution of the circumburst
medium\refto{Chevalier99}.

\refis{Katz94}  Katz, J.I.
                Low-frequency spectra of gamma-ray bursts.
                {\it Astrophys. J.} {\bf 432}, L110--113 (1994).

\refis{MR97}   M\'esz\'aros, P., \& Rees, M.J.
               Optical and Long-Wavelength Afterglow from Gamma-Ray Bursts.
               {\it Astrophys. J.} {\bf 476}, 232--240, (1997).

\refis{Vietri97} Vietri, M.
               The Soft X-Ray Afterglow of Gamma-Ray Bursts, A Stringent
               Test for the Fireball Model.
               {\it Astrophys. J.} {\bf 478}, L9--12. (1997).

\refis{Waxman97a}   
               Waxman, E.
               Gamma-ray-burst afterglow: supporting the cosmological
               fireball model, constraining parameters, and making
               prediction.
               {\it Astrophys. J.} {\bf 485}, L5--L8 (1997).

\refis{Sari98}
        Sari, R., Piran, T., Narayan, R.
        Spectra and Light Curves of Gamma-Ray Burst Afterglows.
        {\it  Astrophys.~J.} {\bf 497}, L17--L20 (1998).

\refis{Meszaros98}
	M\'esz\'aros, P., Rees, M. J. \&\ Wijers, R. A. M. J.
	Viewing Angle and Environment Effects in Gamma-ray Bursts:
	Source of Afterglow.
	{\it Astrophys. J.} {\bf 499}, 301--308 (1998).

\refis{Chevalier99}
	Chevalier, R. A., \&\ Li, Z.--Y.
	Gamma-Ray Burst Environments and Progenitors
	astro-ph/9904417, http://xxx.lanl.gov, (1999).

 From our spectroscopic observations of March 29th (Figure~\Spectrum)
we find $\beta= -0.8\pm 0.4$.  This combination of ($\alpha,\beta$) is
similar to the ($\alpha=-2.05\pm 0.04, \beta=-1.20\pm 0.25$) seen in
GRB 980519 (ref. \Ref{Halpern99}) and can be reasonably
interpreted\refto{Sari99} as arising from a standard $p \sim 2.2$
shock with a jet-like emitting surface. Alternatively, the emission
could arise in a $p \sim 3$ shock propagating in a circumburst
medium\refto{Chevalier99} whose density falls as the inverse square of
the distance from the explosion site.

\refis{Halpern99}
	Halpern, J. P., Kemp, J., Piran, T., \&\ Bershady, M. A.
	The Rapidly Fading Optical Afterglow of GRB 980519.
	astro-ph/9903418, http://xxx.lanl.gov, (1999).

\refis{Sari99}
	Sari, R. \&\ Piran, T.
	Jets in GRBs.
	astro-ph/9903339, http://xxx.lanl.gov (1999).

\medskip
\noindent{\bf A new transient source}
\medskip

We now discuss the bright source seen in the rebrightening phase
(corresponding to observations of April 17 and April 23). This source
is $\sim 60$ times brighter than that extrapolated from the rapidly
declining afterglow. The magnitude of this excess and the late
timescale of rebrightening has never been reported before.

We first establish the reality of the source.  As noted in the legend
to Figure~\Image\ the source is consistently detected in three
separate images of April 17.  In the summed image, the source is
detected at 4.6-$\sigma$ (chance probability of $2\times 10^{-6}$).
We note that all other objects in the field at this flux level are
reliably detected in our deeper December 18th image. Next, the source
is clearly detected in the spectrum obtained on April 23rd
(Figure~\Spectrum) and at the same position as that of the OT.
Finally, we note that the source in the April 17th image is coincident
with the position of the OT in the image of March 27th to within the
expected astrometric error, $0.04\pm 0.18$ arcsecond.

We conclude that there was indeed a source at the position of the OT
which brightened three weeks after the burst and subsequently
faded to undetectable levels. We now investigate possible explanations
for this source.

The simplest picture is that the rebrightening phase is due to a
rebrightening of the optical afterglow itself. As noted earlier, this
would be unprecedented in both the magnitude of the rebrightening and
the epoch of rebrightening.  Piro et al.\refto{Piro99} have recently
suggested that the doubling of the X-ray flux of GRB 970508 three days
after the GRB event arises from the relativistic shell running into a
dense gas cloud.  Such an explanation for the GRB 980326 light curve
would require a large dense region, with a size comparable to the
timescale of rebrightening, about $\Gamma\, \times$ 10 light days
($\sim 0.1$ pc) and located at a distance $\Gamma^2\, c\, \times 20$
days ($\sim 1$ pc) from the explosion site.  Here, $\Gamma$ is the
bulk Lorentz factor of the shock and is expected to be order unity
three weeks after the burst.  Panaitescu et al.\refto{Panaitescu98}
suggest the rebrightening of GRB 970508 may be due to a shock
refreshment -- delayed energy injection by the extremely long-lived
central engine that produced the GRB.  In both these models, the
expected spectrum would be the typical synchrotron spectrum, flux
$f_\nu\propto
\nu^{-1}$ (or flatter). The very red spectrum of April 21st
(Figure~\Spectrum) allows us to essentially rule out a synchrotron
origin for the rebrightening phase of GRB 980326.

\refis{Piro99}
	Piro, L. et al.  
	The X-Ray Afterglow of the Gamma-Ray Burst of
	1997 May 8: Spectral Variability and Possible Evidence of an
	Iron Line.  
	{\it Astrophys. J.} {\bf 514}, L73--L77, (1999).

\refis{Panaitescu98}
	Panaitescu, A., M\'esz\'aros, P., \& Rees, M.J.
	Multiwavelength Afterglows in Gamma-Ray Bursts: Refreshed
	Shock and Jet Effects
	{\it  Astrophys.~J.} {\bf 503}, 314--324 (1998).

Alternatively, the GRB could have occurred in a dusty region and the
afterglow would rebrighten\refto{Loeb99} as the dust is sublimated by
the afterglow.  However, the observed spectral evolution from a
relatively blue spectrum (March 29) to red (April 23) moves in a
direction opposite to that expected in this scenario.

\refis{Loeb99}
	Loeb, A.  Talk given at the Institute for Theoretical Physics,
	University of California, Santa Barbara, March 1999.

\bigskip
\noindent{\bf The supernova interpretation}
\medskip

We advance the hypothesis that the new source is due to an underlying
supernova (SN) revealed only after the afterglow emission has
vanished.  Woosley and collaborators (see ref.
\Ref{MacFayden99,Woosley93} and references therein) have pioneered the
``collapsar'' model in which GRBs arise from the death of massive
stars -- stars which produce black hole remnants rather than neutron
stars.  In this model, the iron core of a massive star collapses to a
black hole and releases up to a few $\times 10^{52}$ erg of kinetic
energy. Some fraction of this energy is expected to emerge in the form
of a jet with little entrained matter; bursts of gamma-rays result
from internal shocks in this jet.  The remaining energy is absorbed by
the star, causing it to explode and thereby produce an energetic
supernova.

\refis{MacFayden99}
	MacFadyen, A. \&\ Woosley, S. E. 
	Collapsars -- Gamma-Ray Bursts and Explosions in 
	``Failed Supernovae''.
	astro-ph/9810274, http://xxx.lanl.gov, (1999).

Thus in this model, the total light curve has two distinct
contributions: a power-law decaying afterglow component and emission
from the underlying supernova.  In Figure~\Light-Curve\ we present the
light curve expected in this model and use the light curve of the well
observed\refto{Galama98,McKenzie99} SN 1998bw as a template for the
supernova contribution.  We find the $R$-band and $I$-band data
consistent with a bright supernova at $z \approx 1$.

\refis{McKenzie99}
	McKenzie, E. H. \&\ Schaefer, B. E.
	The Late Time Light Curve of SN 1998bw Associated with
	GRB980425
        astro-ph/9904397, http://xxx.lanl.gov, (1999).

The very red spectrum of the source on April 23 finds a natural
explanation in the supernova hypothesis.  From
theoretical\refto{MacFayden99} and phenomenological\refto{Kulkarni98b}
grounds we expect GRBs to arise from massive stars which have lost
their hydrogen envelope, i.e.~Type Ibc supernovae. At low redshifts,
all Type I supernovae are observed to exhibit a strong UV deficit
relative to the blackbody fit to their spectra.  This deficit is due
to absorption by prominent atomic resonance lines starting below $\sim
3900$ \AA. Below $\lambda_c \sim 2900$ \AA\ we expect to see very
little flux.  In the near UV range (3000--4000 \AA) type I SNe spectra
have a red appearance. Approximating the flux by a power law ($f_\nu
\propto \nu^\beta$), the power law index (depending on the wavelength range
chosen) is $-3$ or even smaller; see ref. \Ref{Kirshner93} for a UV
spectrum of a type Ia SN.  Fitting the spectrum of Figure~\Spectrum\
to a power law we obtain $\beta=-2.8\pm 0.3$.  In this interpretation
the redshift of the source is $z \sim 1$. A smaller redshift would
lead to a larger $\beta$. A larger redshift would substantially
suppress the light in the observed $R$ band (which covers the
wavelength range 5800--7380 \AA).  Indeed, from Figure 1, we can
deduce that $z \ale 1.6$.

\refis{Kirshner93}
	Kirshner, R. P., et al.
	SN 1992A: Ultraviolet and Optical Studies Based on HST, 
	IUE, and CTIO Observations
	{\it Astrophys. J.} {\bf 415}, 589--615 (1993).

We do not know {\it a priori} the spectrum and light curve of a
supernova accompanying GRBs.  However, we have used the light curve of
SN 1998bw because it is a very well studied Type Ibc SN with a
possible association with GRB 980425.  If SN 1998bw is indeed an
appropriate template for a supernova associated with a GRB then, as
Figure \Light-Curve illustrates, the likely redshift of GRB 980326 is
$z \sim 1$ and, as discussed above, the red spectrum of April 23
similarly suggests $z\sim 1$. Given the low signal-to-noise ratio of
the April 23rd spectrum and the expected line broadening due to high
photospheric velocity we do not, as seems to be the case, expect to
see any spectral features.

Independently, from the absence of strong spectral breaks in our
spectrum of the OT we can firmly place $z_{OT} \ale 2.3$. This
constraint is consistent with our deduction that $z \ale 1.6$ (see
above).  Thus from a variety of accounts we find a plausible redshift
of around unity for GRB 980326. Such a redshift is not entirely
unexpected. Indeed, we note that five out of eight spectroscopically
confirmed redshifts of GRBs lie in the range $0.7 < z < 1.1$.

\bigskip
\noindent{\bf Implications of the supernova connection}
\medskip

The GRBs localized by BeppoSAX belong to a class of long duration
GRBs. The jet in a collapsar model takes many seconds to penetrate the
star, and therefore the collapsar model is unlikely to
account\refto{Woosley98b} for the class of short duration (less than a
few seconds) GRBs.  If we accept the SN interpretation for GRB 980326,
a long duration (5 s) GRB then it is only reasonable to posit that all
other long duration GRBs are also associated with SNe. In what way can
this assertion be tested observationally?

\refis{Woosley98b}
        Woosley, A.  Talk given at Rome Conference on Gamma-Ray
        Bursts, Rome, Italy, November 1998.

The evidence for an underlying SN can come in two ways. First, is the
direct evidence for an accompanying SN seen in the light curve at
timescales comparable to the time for SNe to peak, $\sim 20(1+z)$
days.  However, in our opinion three conditions must be satisfied in
order to see the underlying SN even when one was present.  (1) The GRB
afterglow should decline rapidly, otherwise the SN will remain
overpowered by the afterglow for all epochs.  (2) Given the strong UV
absorption (discussed above), only GRBs with redshift $z \simlt 1.6$
have an observable SN component in the optical band. (3) The host must
be dimmer than the peak magnitude of the SN ($M_V \sim -19.5$). The
last requirement is not needed if the GRB can be resolved from the
host (e.g. with HST). Finally one caveat is worth noting: the peak
magnitudes of Type Ibc SNe are not constant (unlike those of Type Ia)
and can vary from $-16$ mag to a maximum of $-19.5$ mag; see ref.
\Ref{Iwamoto98}.  We have investigated the small sample of GRBs with
adequate long-term follow up and conclude that perhaps only GRB 980519
satisfies the first and the third observational conditions for SN
detection; the redshift of this GRB is unfortunately unknown.

\refis{Iwamoto98}
	Iwamoto, K. et al. 
	A hypernova model for the supernova associated with the
	gamma-ray burst of 25 April 1998.
	{\it Nature} {\bf 395}, 672--674.

The second method is an indirect method to tie GRBs and SNe.  The
dynamics of the relativistic blast wave is strongly affected by the
distribution of circumstellar matter.  Chevalier and
Li\refto{Chevalier99} note that massive stars, through their active
winds, leave a circumstellar medium with density falling as the inverse
square of the distance from the star.  One expects smaller $\alpha$ for
GRBs exploding such a circumstellar medium.  In this framework, GRB
afterglows which decline rapidly and are at modest redshifts will again
be prime targets to search for the underlying SN.

\refis{Chevalier99} 
	Chevalier, R. \&\ Li, Z.-Y.
	Gamma-ray Burst Environments and Progenitors.
	astro-ph/9904417, (1999).

In conclusion we note that it is not possible to firmly demonstrate on
purely observational grounds that all long duration GRBs can be
explained by the collapsar model. However, we strongly urge sensitive
observations especially at longer wavelengths (to avoid the UV cutoff
of SNe) for GRBs satisfying the above three conditions. If the
proposed hypothesis is correct then the light curves and the spectra
would show the behavior shown and discussed in Figure~\Light-Curve\
and Figure~\Spectrum.

We end with a discussion of one interesting point.  The total energy
release in $\gamma$-rays of GRB 980326 was $E_\gamma=3\times
10^{51}f_J$ erg where $f_J$ is the fractional solid angle of the jet
(if any); here we have used the measured fluence\refto{Groot98} and
assumed $z\sim 1$ ($H_0 = 65$ km s$^{-1}$ Mpc$^{-2}$, $\Omega_0 =
0.2$, $\Lambda_0 = 0$).  If this GRB was beamed then $E_\gamma \sim
10^{49}$ erg.  Curiously enough, this rather small energy requirement
places GRB 980326 as close in energetics to GRB 980425 ($E_\gamma = 6
\times 10^{46}$ erg, ref.
\Ref{Galama98}) as to the classic gamma-ray bursts ($E_\gamma \age 6
\times 10^{51}$ erg).



\bigskip
\bigskip

\noindent{\bf Acknowledgment.} 
We thank M.~H.~van Kerkwijk for help with the December 18 observations
at the Keck II telescope and R.~Sari for helpful discussions.  We
gratefully acknowledge the excellent support from the staff at the
Keck Observatory.  The observations reported here were obtained at the
W.~M.~Keck Observatory, made possible by the generous financial
support of the W.~M.~Keck Foundation, which is operated by the
California Association for Research in Astronomy, a scientific
partnership among California Institute of Technology, the University
of California and the National Aeronautics and Space Administration.
SRK's and AVF's research is supported by the National Science
Foundation and NASA.  SGD acknowledges partial support from the
Bressler Foundation.

\vfill\eject
\def\arcsec{\hbox{$^{\prime\prime}$}}
\vskip 1cm
\centerline{\bf Table \TableMagnitudes.
                Keck II Optical Observations$^{\dagger}$ of GRB
                980326$^*$}
\smallskip
\hrule
\smallskip
\settabs\+ 12345678901 & 1234567 & 12345678901 & 123456789012 &
12345678901234567 & 1234567890123& .\cr

\+ Date$^a$
& Band/
& Int.~Time
& Seeing 
& ~~Magnitude$^b$
& Observers \cr

\+  (UT)
& Grating
& (sec)
& (FWHM)
&
& \cr

\smallskip
\hrule
\medskip

\+ Mar 27.35 & ~R & 240 & 0\arcsec .74 & $21.25 \pm 0.03$
	& AVF, DCL, AGR & \cr
\+ Mar 28.25   & ~R & 240 & 0\arcsec .66 & $23.58 \pm 0.07$
	& HS, AD, DS, SAS \cr
\+ Mar 29.27   & 300 & 3600 &    & $24.45 \pm 0.3^{c}$  & HS, AD, DS, SAS \cr
\+ Mar 30.24   & ~R & 900 & 0\arcsec .93 & $24.80 \pm 0.15$
	 & SP, BG, RK, IH \cr
\+ Apr 17.25   & ~R & 900 & 0\arcsec .82 & $25.34 \pm 0.33^d$ 
	& PC, JB \cr
\+ Apr 23.83   & 300 & 5400 &   & $24.9 \pm 0.3^c$ & SGD, SCO \cr 
\+ Dec 18.50   & ~R & 2400 & 0\arcsec .74 & $> 27.3^e$ 
	 & SRK, JSB, MvK \cr
\+ Dec 18.54   & ~I & 2100 & 0\arcsec .74 & $> 25.3^e$ 
	& SRK, JSB, MvK \cr
\+ Mar 24  & ~I & 5450 & 0\arcsec .80 & $> 26.6^e$ 
	& SRK, JSB\cr

\medskip
\hrule
\smallskip

\medskip
\noindent Notes:

\smallskip
\noindent $\dagger$  We used the Keck II 10-m Telescope 2,048 $\times$
2,048 pixel CCD (charged coupled device) Low-Resolution Imaging
Spectrometer\refto{Oke1995} (LRIS) for imaging and spectroscopy of the
GRB field.

\noindent$^*$ The epoch of GRB 980326 is March 26.888, 1998 
(ref. \Ref{Cel98}).

\refis{Oke1995} Oke, J. B., Cohen, J. L., Carr, M., Cromer, J.,
              Dingizian, A., Harris, F. H., Labrecque, S., Lucinio, R.
              Schaal, W., Epps, H. \&\ Miller, J.
              The Keck Low-Resolution Imaging Spectrometer.
              {\it Publ. Astr. Soc. Pacific} {\bf 107}, 375-385
              (1995).

\noindent (a) Mean epoch of the image.  The year is 1998 for all images
	except for that on March 24 for which it is 1999.

\noindent (b) Photometric Calibration. The absolute zero-point of the 
$R$ (effective wavelength\refto{Fukugita95} of $\lambda_{\rm eff}
\approx$ 6588 \AA ) and $I$-bands ($\lambda_{\rm eff}
\approx 8060$ \AA ) were calibrated to the standard Cousins bandpass 
using standard-stars in the field SA98 (ref. \Ref{Landolt92})
and assuming the standard atmospheric correction on Mauna
Kea (0.1 mag and 0.06 mag per unit airmass, respectively).  The
estimated statistical error on the absolute zero-point is 0.01 mag.
We estimate the systematic error (due to lack of inclusion of color
term) to be less than 0.1 mag.
We propagated all photometry to the absolute zero-point derived in the
first epoch of observation using 8 ``secondary'' stars which were
detected with high signal-to-noise ratio, unsaturated, near to the
transient, and common to every epoch; the typical uncertainty in the
zero-point propagation is 0.01 mag.  Thus any systematic error in our
absolute zero-point will not affect the conclusions based on {\it
relative} flux.  The uncertainties quoted in the Table contain all
known sources of error (aperture correction, etc.)  The calibrated
magnitudes of the secondary stars reported in Groot et
al.\refto{Groot98} agree to within the measurement errors.

\noindent (c) Spectrophotometric measurement.  The flux in $\mu$Jy is
determined at 6588 \AA, the central wavelength of the $R_c$ band;
the conversion to magnitude assumes  0 mag equal to 3020 Jy
(ref. \Ref{Fukugita95}). The spectrophotometric magnitudes
are relative to a bright star that was on the slit (for which
we have obtained independent photometry from our images).

\noindent (d) Photometry of the faint source.  Since the 
transient was not detected to significantly fainter levels in later
epochs it is safe to assume that the April 17 detection was that of a
point-source (and not an extended galaxy as we had
earlier\refto{gcn57} believed).  To maximize the signal-to-noise ratio
we choose to measure the photometry in an aperture radius equal to the
FWHM of the seeing and correct for the missing flux outside the
aperture by using the radial flux profiles of bright isolated stars in
the image.  The determination of the optimum sky level (from which we
subtract the total flux in the aperture) is not well-defined.  We
estimate the systematic uncertainty introduced by the uncertainty in
the sky level as 0.25 mag.  The statistical uncertainty (weighted mean
over different background determinations) of the flux was 0.22 mag.
Thus we quote the quadrature sum of the statistical and systematic
uncertainty of 0.33 mag.

\noindent (e) Upper-limits.  On 1998 Dec 18 UT and 1999 March 24 UT
there was no detectable flux above the background at the position of
the optical transient.  We centered 1000 apertures randomly in our
image (approximately $1800 \times 2048$ pixels in size) and performed
weighted aperture photometry with a local determination of sky
background and recorded the counts (``DN'') above background at each
location. The flux contribution from an individual pixel, some radius
$r$ from the center of the aperture, to the total flux was weighted by
a Gaussian with a radial width FWHM equal to the seeing. A histogram
of the resulting flux was constructed. This histogram was decomposed
into two components---a Gaussian with median near zero DN and a long
tail of positive DN corresponding to actual source detections.  We fit
a Gaussian to the zero-median component, iteratively rejecting outlier
aperture fluxes.  Based on the photometric zero-point and using
isolated point sources in the image for aperture corrections, we
computed the relationship between DN within the weighted aperture and
the total magnitude.  In the Table we quote an upper limit
(95\%-confidence level corresponding to 2-$\sigma$ of the Gaussian
fit) at the position of the optical transient.

\refis{Landolt92}
        Landolt, A.
        UBVRI photometric standard stars in the magnitude range
        11.5--16.0 around the celestial equator.
        {\it Astron.~J.} {\bf 104}, 340--376 (1992).

\vfill \eject


\noindent{\bf Figure \Image.}

Images of the field of GRB 980326 at three epochs.  Each images shows
a 54\arcsec $\times$ 54\arcsec\ region centered on the optical
transient (labeled ``OT'').  In all the images, the local background
has been subtracted by a median filter and the resulting image
smoothed (with a two-dimensional Gaussian with $\sigma =0\arcsec.23$).
An unrelated faint source ``f'' in the field is noted for comparison
of the relative limiting flux between the three epochs: it is
marginally detected (at the $\sim 2$-$\sigma$ level) on March 27 and
April 17 but well detected on December 18.  In contrast the OT is
brighter and better detected (at the 4.6-$\sigma$ level, see text) on
April 17 but clearly not detected to fainter levels on December 18 ($R
> 27.3$; see Table 1).

\noindent{\it Methodology.} 
In keeping with standard practice, our April 17th observations
consisted of three separate 300-s observations (dithered by 5
arcseconds).  Visual inspection of the three frames reveals a faint
source near the position of the optical transient.  In no frames was
did a diffraction spike of the nearby bright star ``A'' overlap the OT
position.  Also, there were no apparent cosmic-ray hits at the
transient position nor were there any strong gain variations (i.e.~no
apparent problem with the flat-fielding) at the three positions on the
CCD.  In both the sum and mode-scaled median of the three shifted
images, we detect a faint source consistent with the centroid location
(angular offset $0.04 \pm 0.18$ arcsec) of the optical transient on
1998 March 27.  Lastly, we computed the point source sensitivity in
the April 17th image by computing Gaussian-weighted photometry in 1000
random apertures (see discussion accompanying Table 1).  Relative to
this distribution, the flux at the location of the transient is
positive and equal to 4.6-$\sigma$; the probability that the measured
flux is due to noise is $2\times 10^{-6}$.  All objects at the flux
level of the transient are reliably detected in the deeper image from
December, thereby providing an independent validation of our
methodology.  We conclude that indeed the transient was significantly
detected on April 17th.  We discuss the photometric calibration of the
detection in Table \TableMagnitudes.

\centerline{\psfig{file=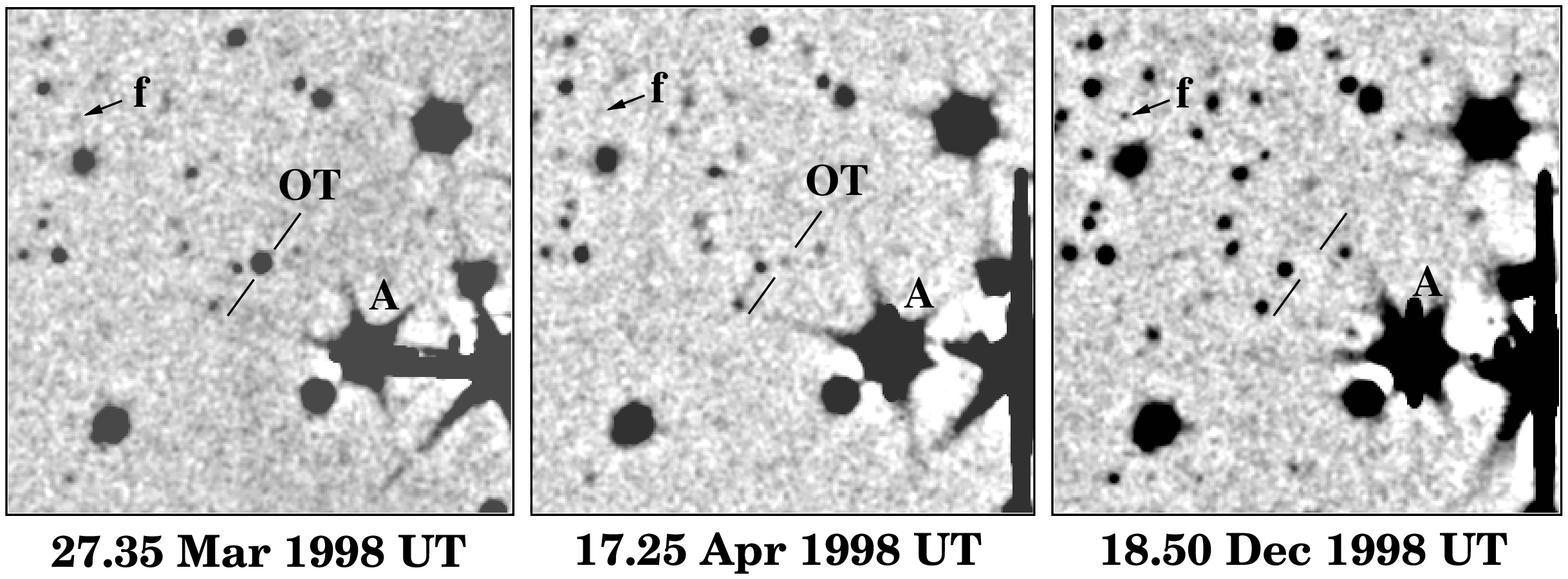,width=6.5in,angle=270}}

\vfill\eject

\noindent{\bf Figure  \Light-Curve.}

The $R$-band light curve of the afterglow of GRB 980326.  Overlaid is
a power-law afterglow decline summed with a bright supernova light
curve at different redshifts.  (Although we use as as a template the
multi-band light curve of SN 1998bw (refs.
\Ref{Galama98,McKenzie99}), the
bright supernova potentially associated with GRB 980425, we emphasize
that the exact light-curve shapes of a supernova accompanying a GRB is
not known $a~priori$.)  The GRB+SN model at redshift of about unity
provides an adequate description of the data.

\noindent
{\bf Transient Light curve.} From ref.~\Ref{Schlegel98} we estimate
the Galactic extinction in the direction of the optical transient
$(l,b = 242{^\circ}.36, 13^{\circ}.04)$ to be E(B$-$V) = 0.08. Thus,
assuming the average Galactic extinction curve ($R_V$ = 3.1), the
extinction measure is $A_R = 0.22$, $A_I$ = 0.16 mag.  Plotted are the
extinction corrected magnitudes (see Table 1) of the transient
converted to the standard flux zero-point of the Cousins $R$ filter
from ref.~\Ref{Fukugita95}). In addition to our data, we include
photometric detections from Groot et al.\refto{Groot98} and an
upper-limit from Valdes et al.\refto{Valdes98} (KPNO).  The GRB
transient flux dominates at early times, but with a power-law decline
slope $\alpha = -2$ (straight solid line).

\refis{Fukugita95}
	Fukugita, M., Shimasaku, K., \&\ Ichikawa, T.
	Galaxy Colors in Various Photometric Band Systems.
	{\it Publ. Astr. Soc. Pacific} {\bf 107}, 945--958 (1995).

\noindent
{\bf Supernova light curve.} The supernova light curve template was
constructed by spline-fitting the broadband spectrum measured by
Galama et al.\refto{Galama98} of the bright supernova 1998bw at
various epochs (augmented with late-time observations of SN 1998bw by
McKenzie and Schaefer\refto{McKenzie99}) and transforming back to the
restframe of SN 1998bw ($z=0.0088$).  As discussed in the text, we
expect the rest-frame UV emission (below 3900\AA) to be suppressed due
to absorption by resonance lines.  We assume that the UV flux declines
as $f_\nu\propto\nu^{-3}$.  Theoretical light curves are then
constructed by red-shifting the template to various redshifts and
determining the flux in the $R$ (observer frame) by interpolating (or
extrapolating, for $z \age 1$).  The flux normalization of the
redshifted SN 1998bw curves are independent of the Hubble constant but
are dependent upon the value of $\Omega_0$ and $\Lambda_0$ (here we
show the curves for $\Omega_0 = 0.2$ and $\Lambda_0=0$).  Beyond $z
\approx 1.3$ the observed $R$-band corresponds to restframe $\lambda
\ale 2900$ \AA. As stated in the text, the spectrum in this range has
been modeled along simple lines.  Qualitatively, the peak flux derived
from the SN model as a function of redshift and shown here agrees with
the theoretical peak flux-redshift relation for type Type
Ia\refto{Schmidt98}. This suggests that our adopted model for the UV
spectrum is reasonable.

\refis{Schlegel98}
        Schlegel, D. J., Finkbeiner, D. P., \& Davis, M.
        Maps of Dust Infrared Emission for Use in Estimation of
        Reddening and Cosmic Microwave Background Radiation Foregrounds.
        {\it Astrophys.~J.}, {\bf 500}, 525--553 (1998). 

\refis{Galama98}
	Galama, T. J. et al.
	An unusual supernova in the error box of the gamma-ray burst of
        25 April 1998.
	{\it Nature}, {\bf 395}, 670--672 (1998).

\refis{Schmidt98}
	Schmidt, B. P. et al.
	The High-$z$ supernova search: Measuring cosmic deceleration
	and global curvature of the universe using type Ia supernovae.
	{\it Astrophys.~J.} {\bf 507}, 46--63 (1998).

\medskip
\centerline{\psfig{file=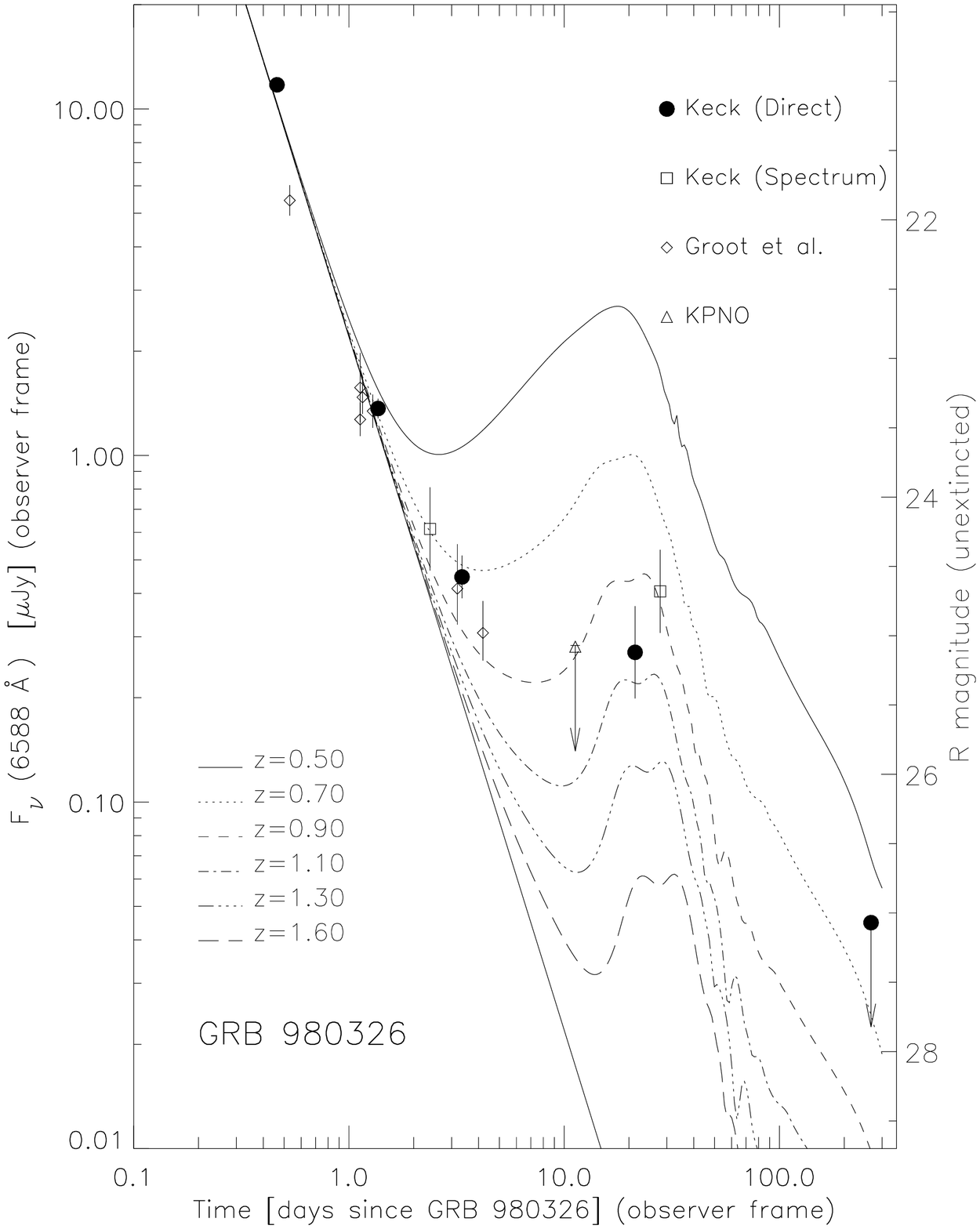,width=5in}}

\vfill\eject

\noindent{\bf Figure~\Spectrum}

The spectra of the transient on March 29.27 and April 23.83 1998 UT.
The two spectra are shown at two different spectral resolutions.
Starting from the top, panels 1 and 3 show the spectra at the two
epochs at the full spectral resolution (see below for details) and
panels 2 and 4 show the same two spectra but binned in groups of 51
channels. Panel 5 is the spectrum of the sky.  The best-fit power law
models ($f_\nu\propto \nu^{\beta}$) to the binned spectra are shown by
dashed lines; the fits were restricted the wavelength range 4500-8500
\AA. The scatter of individual channel values within each bin was used
to assign relative weights to the median fluxes in each bin when
performing the fits.

\noindent{\bf Results.}
On March 29.27, we obtain $\beta= -0.8\pm 0.4$ and $\beta=-2.8\pm 0.3$
on April 23.83.  The derived power-law indices include the correction
of Galactic extinction.  From the absence of continuum breaks in the
spectrum of March 29, we can place an upper limit to the redshift,
$z_{OT} \ale 2.3$.

\noindent
{\bf Observing Details.} Spectroscopic observations of the OT were
obtained on 29 March 1998 UT, using the Low Resolution Imaging
Spectrometer (LRIS)\refto{Oke1995} at the Keck-II 10-m telescope on
Mauna Kea, Hawaii.  We used a grating with 300 lines mm$^{-1}$ blazed
at $\lambda_{\rm blaze} \approx 5000$ \AA\ and a 1.0 arcsec wide slit.
The effective wavelength coverage was $\lambda \sim 4000 - 9000$
\AA\ and the instrumental resolution was $\sim 12$ \AA.  Two exposures
of 1800 s each were obtained.  We used Feige 34 (ref. \Ref{Massey88})
for flux calibration.  The estimated uncertainty of the flux zero point
is about 20\%.  Additional spectra were obtained on 23 April 1998 UT,
in photometric conditions, using the same instrument, except that the
spectrograph slit was 1.5 arcsec wide. The effective spectral
resolution for this observations was $\sim$ 16 \AA.  Three exposures of
1800 s each were obtained.
For these observations we used HD 84937 (ref. \Ref{OkeGunn83}) for
flux calibration. The estimated zero-point uncertainty is  about 10\%.
On both epochs, exposures of arc lamps were used for primary wavelength
calibration. Night sky lines were used to correct for calibration
changes due to flexure.  In both cases, slit position angles were close
to be close to the parallactic angles. Thus the differential slit
losses were negligible.

	\refis{Massey88}
	Massey, P., Strobel, K., Barnes, J. V., Anderson, E.
        Spectrophotometric standards
	{\it Astrophys.~J.} {\bf 328}, 315--333 (1988).

	\refis{OkeGunn83}
	Oke, J. B., \&\ Gunn, J. E.
	Secondary standard stars for absolute spectrophotometry 
	{\it Astrophys.~J.} {\bf 266}, 713--717 (1983).

The spectra shown were convolved with a Gaussian with $\sigma = 5$
\AA\ (i.e., less than the instrumental resolution) and rebinned to a
common 5 \AA\ sampling.  None of the apparent features in the spectra
are real, on the basis of a careful examination of two-dimensional,
sky-subtracted spectroscopic images: apparent emission of absorption
features are all due to an imperfect sky subtraction noise.  A sky
spectrum from the April 23 observation, extracted in the same
aperture, is shown for the comparison.  These spectra are shown before
the correction for the Galactic foreground extinction.

\noindent
{\bf Spectrophometry.}
In both epoch, we chose a slit position angle close to parallactic so
that the slit would cover both the transient and a relatively bright
star ($R \sim 19$).  The spectroscopic $R$-band magnitudes reported in
table 1 were derived relative to the calibrated $R$-band magnitude of
these stars. This calibration serves to eliminate most of the
systematics and calibration errors; that is, the spectrophotometric
magnitudes were put on the direct CCD system, and are not based on the
flux calibration of the spectra (which do, nevertheless, agree to
twenty percent). This procedure bypasses most of the systematic
errors in comparing our spectroscopic magnitudes with those
from direct CCD images.

\centerline{\psfig{file=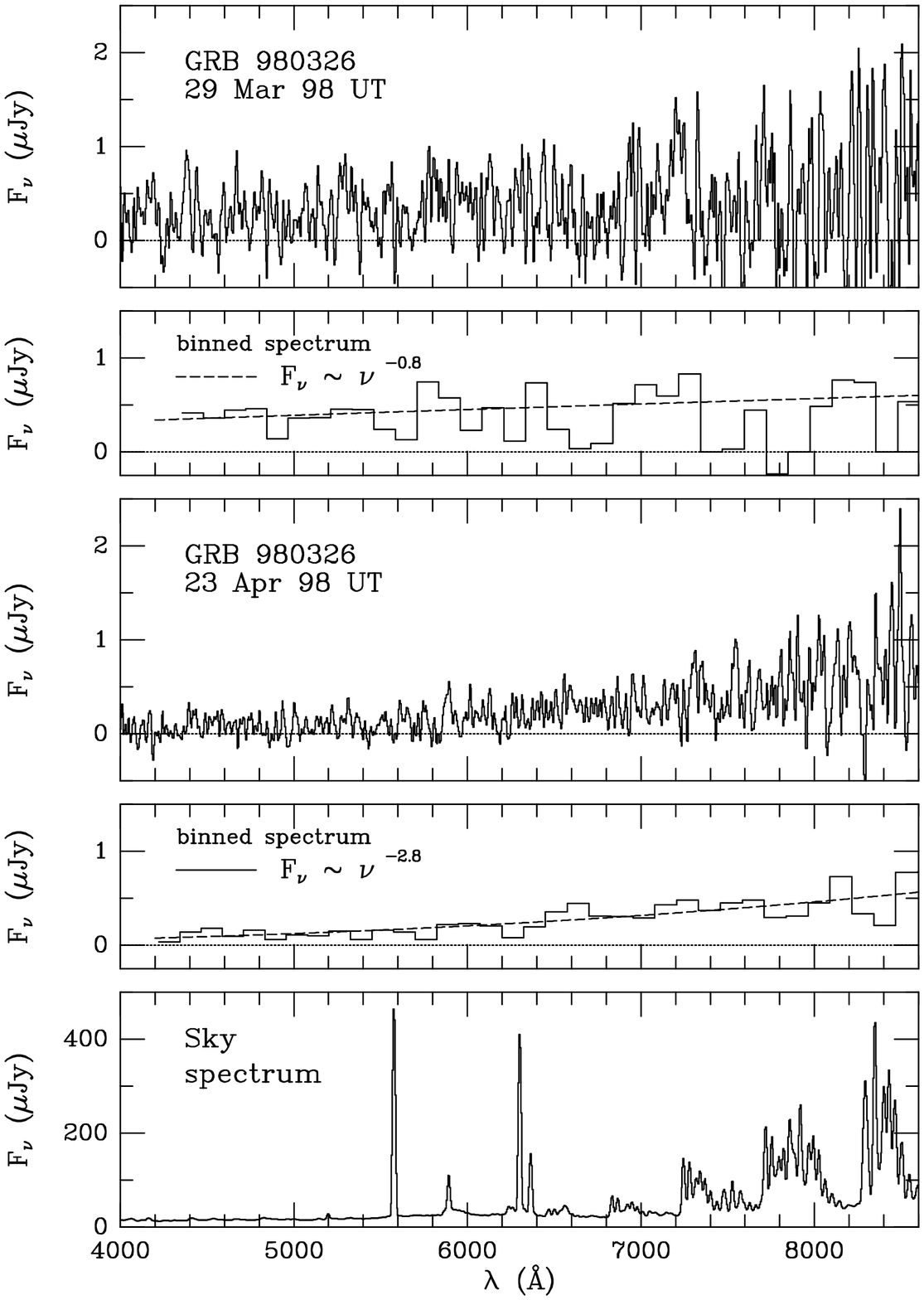,width=5in,angle=0}}

\vfill\eject
\centerline {\bf References}
\bigskip

\endreferences

\vfill\eject

\endmode
\bye